\documentclass[aps,prd,twocolumn,a4paper,10pt]{revtex4}
\usepackage{blindtext}
\usepackage{graphicx}
\usepackage{amssymb,amsmath,amscd}
\usepackage{lipsum}
\usepackage{physics}
\usepackage{mathrsfs}
\usepackage{MnSymbol}
\usepackage[colorlinks=true, linkcolor=blue, citecolor=blue]{hyperref}

\DeclareMathOperator{\diag}{diag}

\newcommand{\be}{\begin{equation}}
\newcommand{\ee}{\end{equation}}

\newcommand{\PR}[1]{\ensuremath{\left[#1\right]}}
\newcommand{\PC}[1]{\ensuremath{\left(#1\right)}}
\newcommand{\chav}[1]{\ensuremath{\left\{#1\right\}}}

\begin{document}
\title{Exact solutions for a complex scalar field under discrete  symmetry}

\author{D. Bazeia$^1$
\!\!\footnote{Corresponding author. Email: dbazeia@gmail.com}}
\author{R. Menezes$^2$
\!\!\footnote{Email: betomenezes@gmail.com}}
\author{G. S. Santiago$^1$
\!\!\footnote{Email: gss.santiago99@gmail.com}}
\affiliation{${^1}\!\!$ {Departamento de Física, Universidade Federal da Paraíba, João Pessoa, Paraíba, Brazil}}
\affiliation{${^2}\!\!$ {Departamento de Ciências Exatas, Universidade Federal da Paraíba, Rio Tinto, Paraíba, Brazil}}

\begin{abstract}
We report on the presence of families of exact solutions for a complex scalar field that behaves according to the rules of discrete $Z_N$ symmetry. Since the family of models is exactly solved, the results appear to be of interest to integrability, to build junctions and networks of localized structures and to describe scalar dark matter in high energy physics.
\end{abstract}

\maketitle

In this Letter, we investigate the presence of exact solutions in models described by a single complex scalar field in $1+1$ spacetime dimensions. The study focuses on models described by a complex scalar field engendering the $Z_N$ symmetry, and is motivated by other investigations, as the ones described in  \cite{F,AT,B1,B2,orb1,orb2,orb3} and in references therein.

The interest concerns distinct possibilities, for example, the use of the deformation procedure previously introduced in \cite{DP}, which can deform a given model described by a potential containing some minima, to get to another model, with the potential giving rise to another set of minima, which may increase periodically. This case was previously considered in Ref. \cite{orb2}, and strongly suggests the application of the deformation procedure to build junctions and networks of localized structures. Evidently, the construction of networks has wider interest and may be considered to describe crystalline behavior similar to the case of skrymion crystals; see Refs. \cite{rec20,rec2,rec21} and references therein for further information. 

Another line of investigation of current interest 
relies on the fact that the model to be studied below may have direct connection to integrability and dark matter. Since the $Z_N$ group is the center of $SU(N)$, our results may be of direct interest to the recent study \cite{DM1}, in which scalar fields charged under a $SU(N)$ gauge group have been used to show that its center, the discrete subgroup $Z_N$, can contribute to ensuring the stability of scalar dark matter particles.

We aim to explore models involving a single complex scalar field, searching for solutions that connect the vertices of regular polygons inscribed in the unit circle on the complex plane. We introduce an \textit{ansatz} that ensures the solutions remain confined within the unit circle and interpolate continuously between the vertices of the regular polygons. All fields, parameters and spacetime coordinates are rescaled to be dimensionless and we consider a $1+1$ dimensional Minkowski spacetime with metric signature $\eta^{\mu\nu}=\diag\PC{1,-1}$. The study starts by supposing that the system is controlled by the following Lagrangian density 
\begin{equation}
\label{L}
    \mathcal{L} = \frac{1}{2}\partial_{\mu}\varphi\partial^{\mu}\Bar{\varphi} - V(\varphi,\overline{\varphi}),
\end{equation}
where $\varphi$ is a complex scalar field and $\Bar{\varphi}$ stand for its complex conjugate. The potential $V(\varphi,\overline{\varphi})$ is a nonnegative real function that modulates the field self-interactions
and is the source of nonlinearities, responsible for the presence of nontrivial solutions in the model. Given that the potential is nonnegative, we can define it as follows
\begin{equation}
    V(\varphi,\overline{\varphi})=\frac{1}{2}\abs{W'(\varphi)}^2,
\end{equation}
where the auxiliary function $W(\varphi)$ is supposed to be holomorphic, with $W'(\varphi) = dW/d\varphi$. The global minima of the potential correspond to the critical values $\chav{v_n}$ of the function $W^{\prime}(\varphi)$, i.e., $W^{\prime}(v_n) =0$. It is also of interest to highlight that $W(\varphi)$ is defined up to a phase factor $e^{i\xi}\in U(1)$, which leaves the Lagrangian density invariant. So, it is possible to define
\begin{equation}
    W_{\xi}(\varphi) = e^{i\xi}W(\varphi).
\end{equation}
Also, to support nontrivial solutions, the vacuum manifold must be degenerate, i.e., there must exist at least two disconnected global minima in the potential.

For the model defined in \eqref{L}, the corresponding equations of motion are
\begin{subequations}
\label{EoM}
    \begin{align}
\frac{1}{2}\partial_{\mu}\partial^{\mu}\varphi + V_{\overline{\varphi}}&=0,\\
\frac{1}{2}\partial_{\mu}\partial^{\mu}\Bar{\varphi} + V_{\varphi}&=0.
    \end{align}
\end{subequations}
Moreover, since the system engenders translational invariance, we are able to define a conserved energy-momentum tensor, which acquires the form
\begin{equation}
    T^{\mu\nu} = \frac{1}{2}\PC{\partial^{\mu}\varphi\partial^{\nu}\Bar{\varphi} + \partial^{\mu}\Bar{\varphi}\partial^{\nu}\varphi} - \eta^{\mu\nu}\mathcal{L}.
\end{equation}
This leads to the energy density which can be expressed as
\begin{equation}
\label{rho}
    \rho = \frac{1}{2}\PC{\abs{\Dot{\varphi}}^2+\abs{\varphi^{\prime}}^2} + \frac{1}{2}\abs{W'(\varphi)}^2.
\end{equation}

We now deal with static configurations, with $\Dot{\varphi}=0$. To obtain spatially localized solutions, each contribution to the energy density must vanish independently in the asymptotic limits. As a consequence, one has to impose as boundary conditions that:  $\varphi^{\prime}\vert_{x\rightarrow\pm\infty}\rightarrow0$ and $\varphi\vert_{x\rightarrow\pm\infty}\rightarrow v_\pm$, where $W^{\prime}(v_\pm)=0$.
Following the Bogomol'nyi procedure \cite{bogo}, we construct a first-order framework that ensures the minimization of the energy, resulting in stable solutions. To implement this procedure, we rewrite the energy density \eqref{rho} to obtain
\begin{equation}
        \rho =  \frac{1}{2}\vert \varphi^{\prime}-\overline{W_{\xi}^{\,\prime}} \vert^2+\operatorname{Re}\PC{\frac{dW_{\xi}}{dx}}.
        \end{equation}
This implies on a energy lower bound
\begin{equation}
        E \geq E_B = \int dx\operatorname{Re}\PC{\frac{dW_{\xi}}{dx}},
\end{equation}
which is saturated when we ensure that
\begin{subequations}
    \begin{align}
        \varphi^{\prime} &= \overline{W_{\xi}^{\,\prime}(\varphi)},\label{1o}\\
        \overline{\varphi}^{\,\prime} &= W_{\xi}^{\prime}(\varphi)\label{1oc}.
    \end{align}
\end{subequations}
Under these conditions, the energy reduces to
\begin{equation}
    E_B = \vert\operatorname{Re}\PR{W_{\xi}(v_+)-W_{\xi}(v_-)}\vert,
\end{equation}
which is completely determined by the boundary conditions on the scalar field. As it was previously demonstrated in Refs. \cite{B1,B2}, since $W$ is holomorphic, all the solutions of the equations of motion are also solutions of the first-order equations, so they are all BPS solutions \cite{bogo,PS}. Moreover, the pair of BPS equations (\ref{1o},\ref{1oc}) must satisfy the orbit relation \cite{orb1,orb2,orb3}
\begin{equation}
    d\PC{W_{\xi}(\varphi) - \overline{W_{\xi}(\varphi)}} = 0,
\end{equation}
which reduces to
\begin{equation}
\operatorname{Im}\PR{W_{\xi}(\varphi)} = C,
\end{equation}
where $C$ is a real constant. Since a kink-like solution interpolates between two adjacent minima $v_{\pm}$, we have that
\begin{equation}
\operatorname{Im}\PR{W_{\xi}(v_+) - W_{\xi}(v_-)} = 0.
\end{equation}
This condition can be rearranged as
\begin{equation}
\label{fp}
    \xi = -\arctan\PR{\frac{\operatorname{Im}\PC{W(v_+)-W(v_-)}}{\operatorname{Re}\PC{W(v_+)-W(v_-)}}}, \quad \mod \pi.
\end{equation}
Although $\xi$ is a symmetry parameter of the theory, the system of BPS equations selects a straight-line orbit on the W-plane, which fixes a specific value for $\xi$ for each topological sector.

As an illustration of interest, let us consider the following function
\begin{equation}
\label{W}
        W(\varphi) = \frac{\alpha i}{N}\PC{\varphi^{-N} + 2N\ln(\varphi) - \varphi^{N}},
        \end{equation}
where $N\in\mathbb{N}$ and $\alpha$ is a positive real parameter. It implies that
\begin{equation}
        W^{\prime}(\varphi) = -\alpha i \frac{\PC{1-\varphi^N}^2}{\varphi^{N+1}}.
\end{equation}
 The critical points of this function are the $N$-th roots of unity,
\begin{equation}
\label{min}
    v_n = \exp\PC{i\frac{2\pi n}{N}}, \qquad n=0,1,\dots,N-1.
\end{equation}
These points represent the vertices of the regular polygons inscribed on the unit circle. Taking $v_+=v_{n}$ and $v_-=v_{n+1}$, we have that
\begin{equation}
    \Delta W = W(v_{n}) - W(v_{n+1}) =\frac{4\alpha\pi}{N},
\end{equation}
which is real. Therefore, substituting this expression into Eq. \eqref{fp} we find that $\xi=0$. So, the energy of such solution is simply $E_{B} = \vert\Delta W \vert$.

\begin{figure}[!ht]
    \centering
    \includegraphics[scale=0.20]{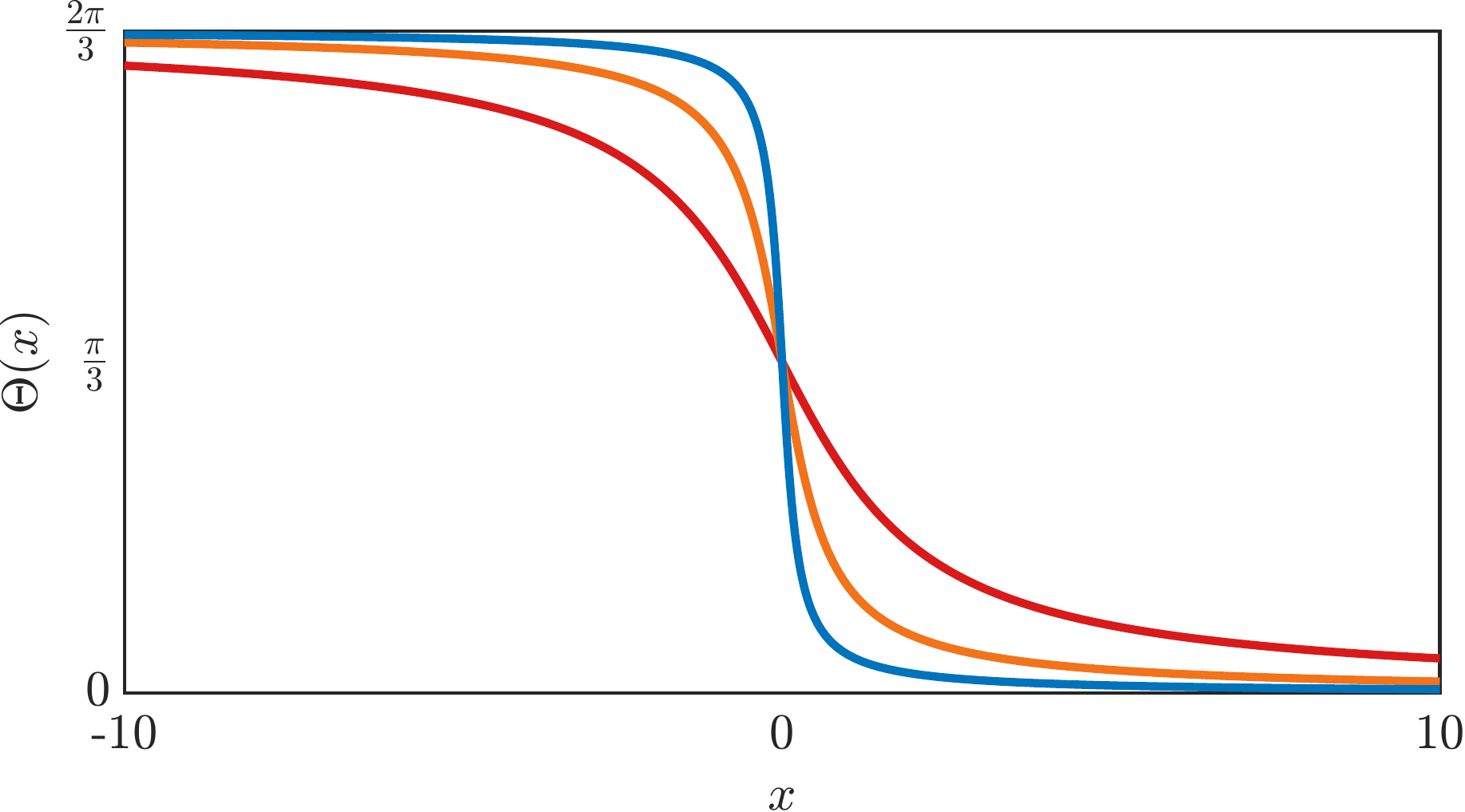}\\
    \includegraphics[scale=0.20]{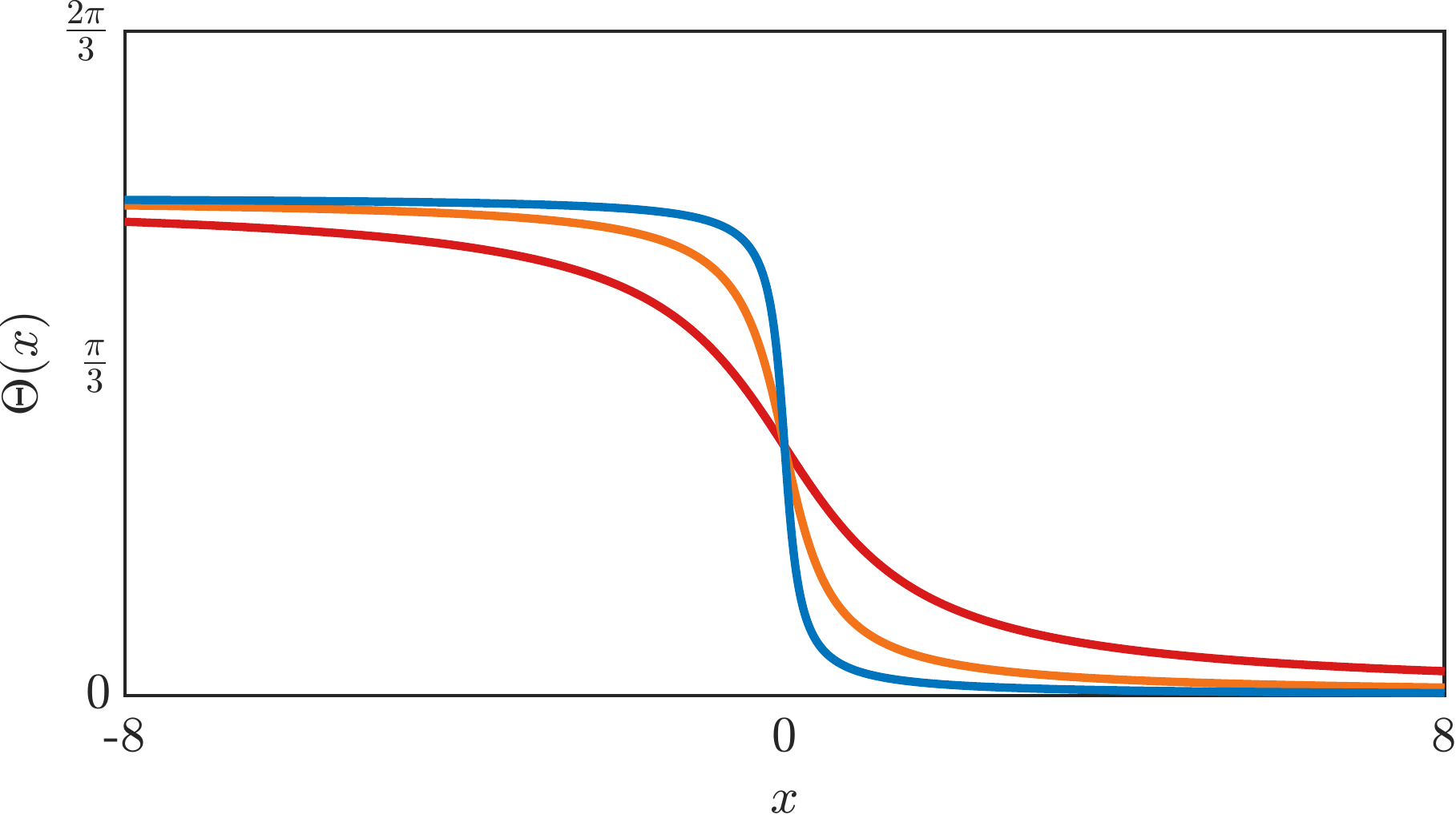}\\
    \includegraphics[scale=0.20]{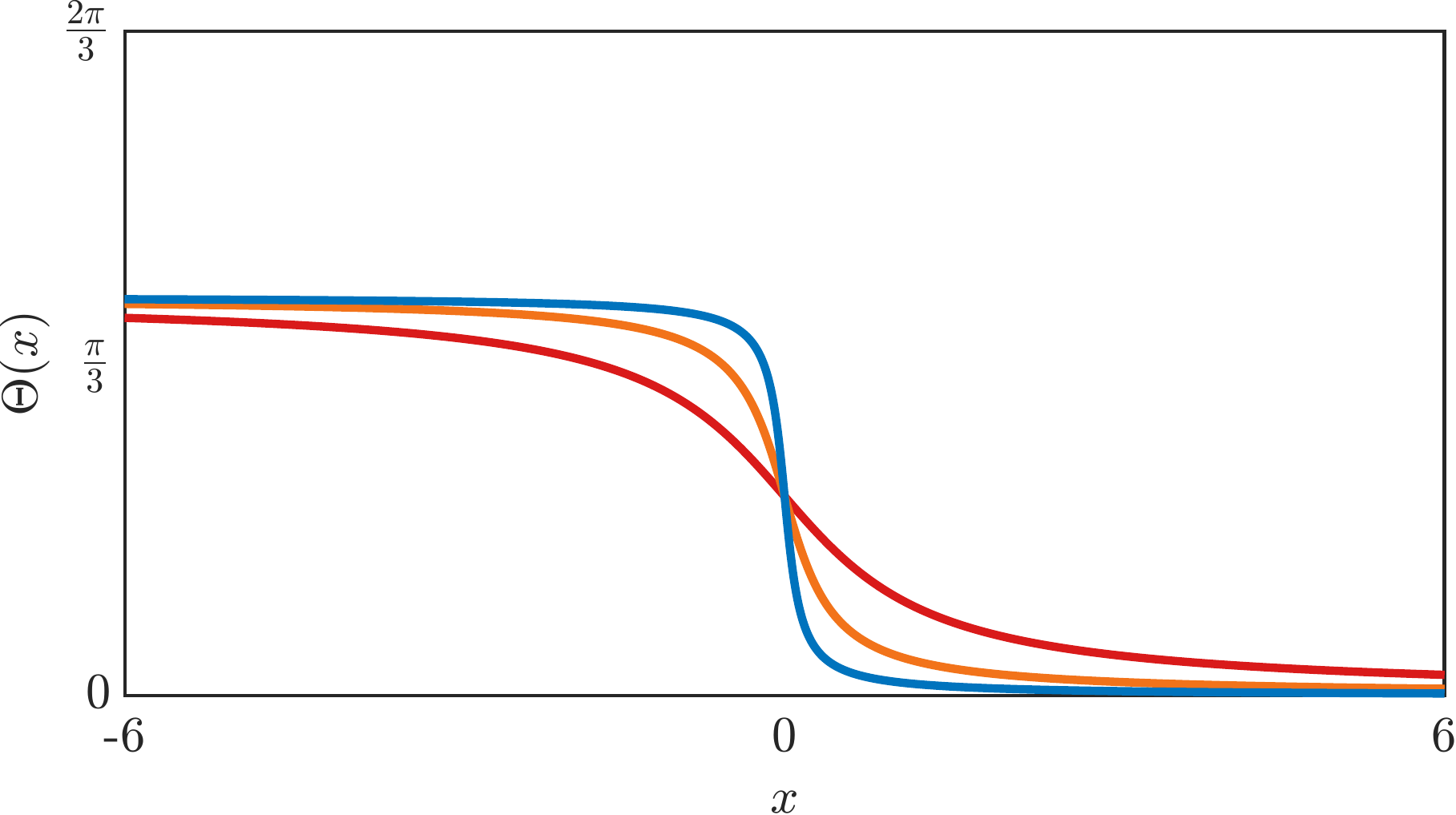}\\
    \includegraphics[scale=0.20]{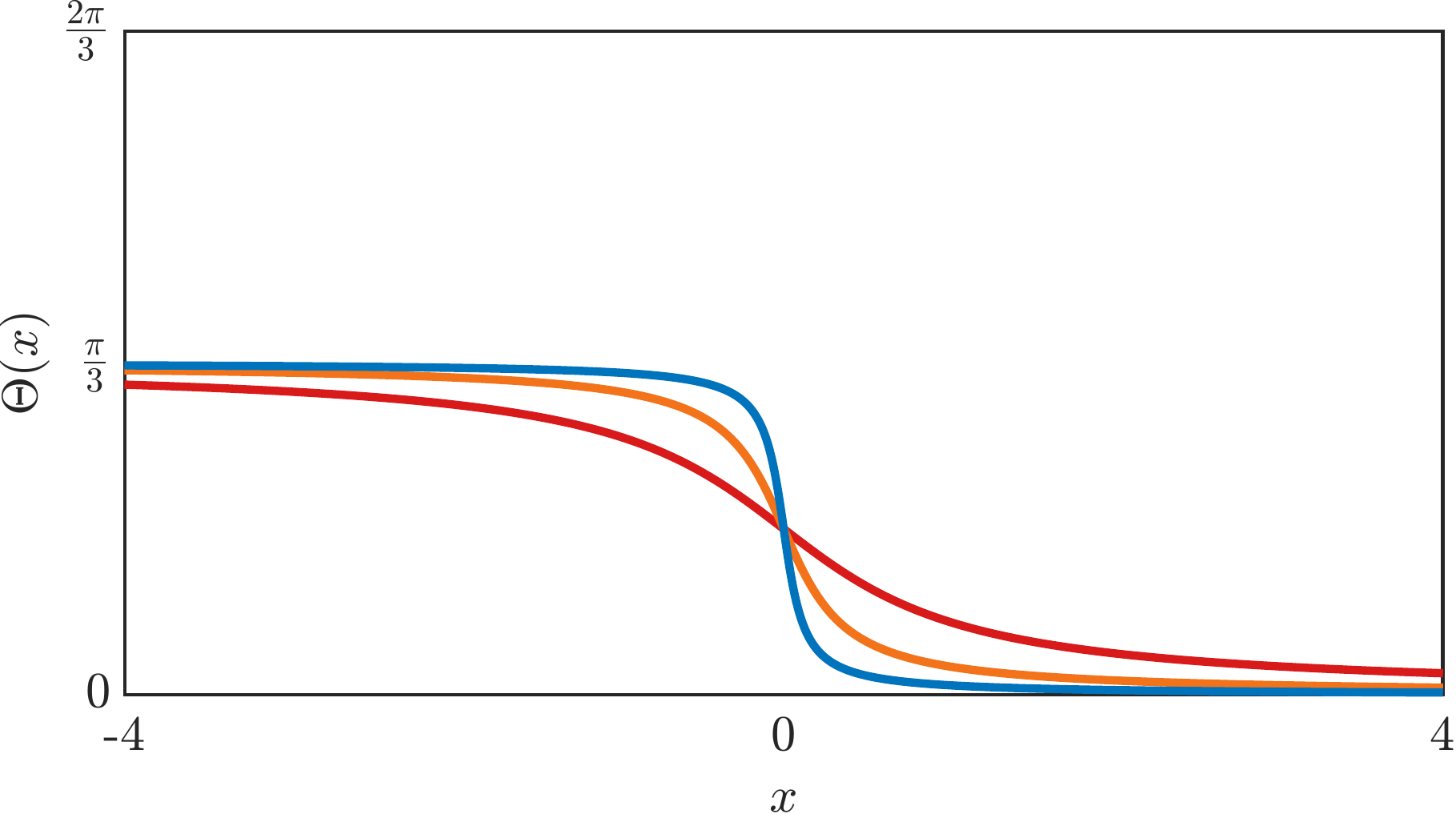}
    \caption{The phase $\Theta(x)$ \eqref{theta} for $N=3$ (top), $4$ (middle top), $5$ (middle bottom) and $6$ (bottom), with $n=0$, $x_0=0$ and $\alpha=0.1$ (red), $0.3$ (orange) $0.9$ (blue).}
    \label{fig1}
\end{figure}

We seek static solutions that connect distinct vacua and remain on the unit circle. Hence, consider that the solution of the complex scalar field $\varphi(x)$ is of the form
\begin{equation}
\label{ansatz}
    \varphi(x) = e^{i\Theta(x)},
\end{equation}
where $\Theta(x)$ is a real function. For this choice to be in accordance with the boundary conditions for localized solutions, we must impose that $\Theta^{\prime}(\pm\infty)=0$, with $\Theta(\pm\infty)$ approaching constant values, corresponding to two neighboring minima of the potential. Substituting this \textit{ansatz} into the first-order equation \eqref{1oc} and simplifying, we find
\begin{equation}
\Theta'=-4\alpha\PC{\sin^2\PC{\frac{N}{2}\Theta}}.
\end{equation}
It can be solved analytically as
\begin{equation}
\label{theta}
    \Theta(x) = \frac{2}{N}\arccot\PC{2\alpha N\PC{x-x_0}} + \frac{2n\pi}{N},
\end{equation}
where $x_0$ is a constant of integration that localizes the center of the function $\Theta(x)$ and $n\in\chav{0,1,\dots,N-1}$ labels the sector. See Fig. \ref{fig1}, where $\Theta(x)$ is depicted for $N=3, 4, 5$ and $6$, and for some values of $\alpha$. Here, we are considering $\arccot$ defined over $\PC{0,\pi}$, so $\Theta(x)$ is a monotonic continuous function which ensures that the solution $\varphi(x)$ interpolates between two adjacent minima of the potential. It has the shape of a kink-like configuration, with the amplitude and width being inversely dependent on $N$ and $\alpha N$, respectively.  

\begin{figure}[!ht]
    \centering
    \includegraphics[scale=0.18]{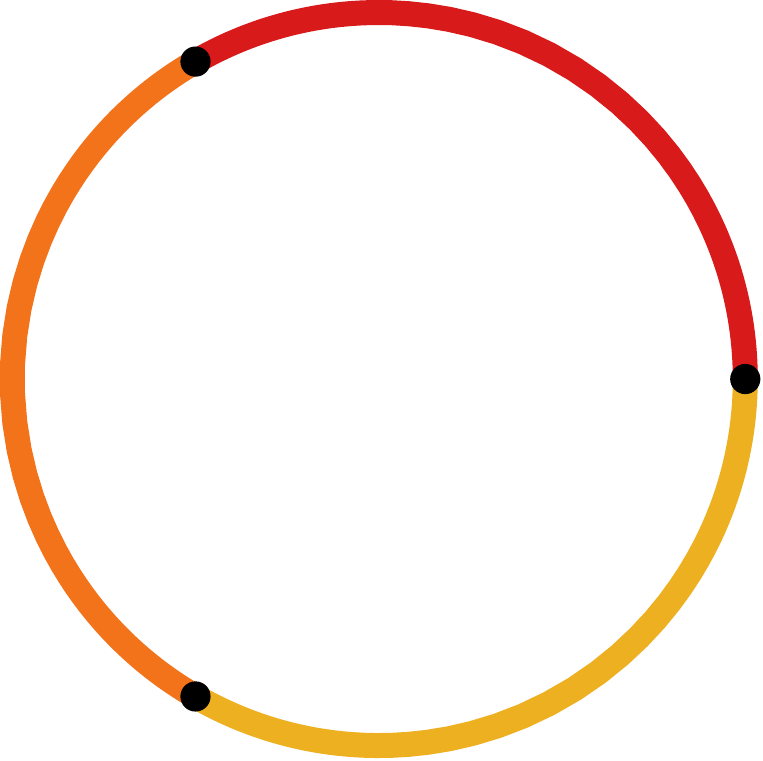}\qquad\quad
    \includegraphics[scale=0.18]{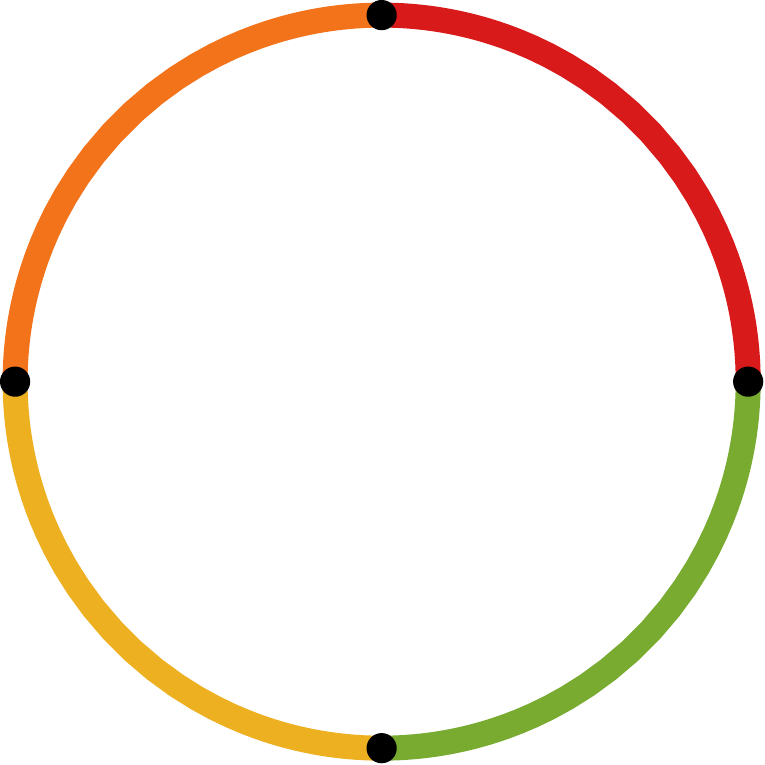}\\ \vspace{12pt}
    \includegraphics[scale=0.18]{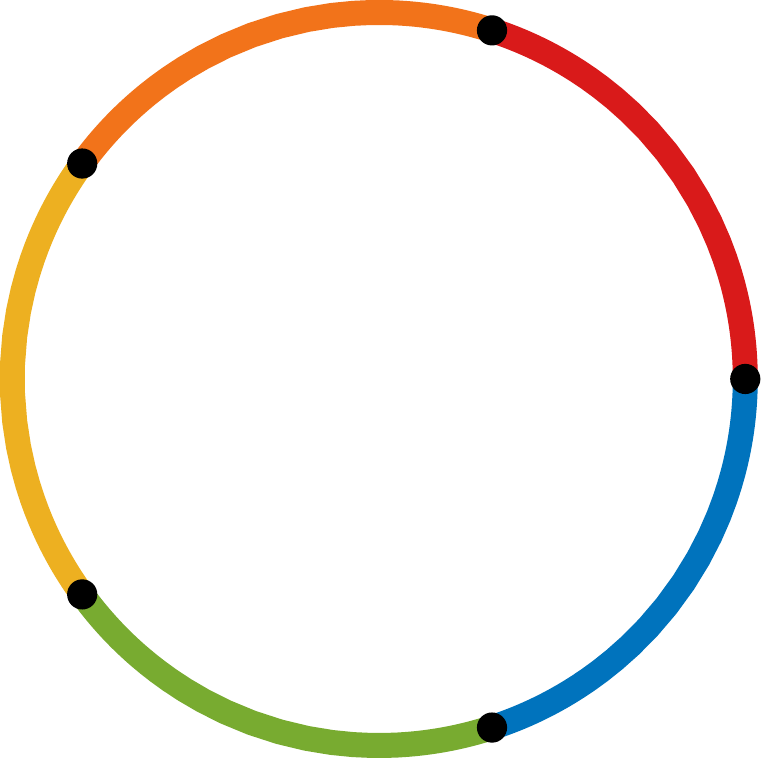}\qquad\quad
    \includegraphics[scale=0.18]{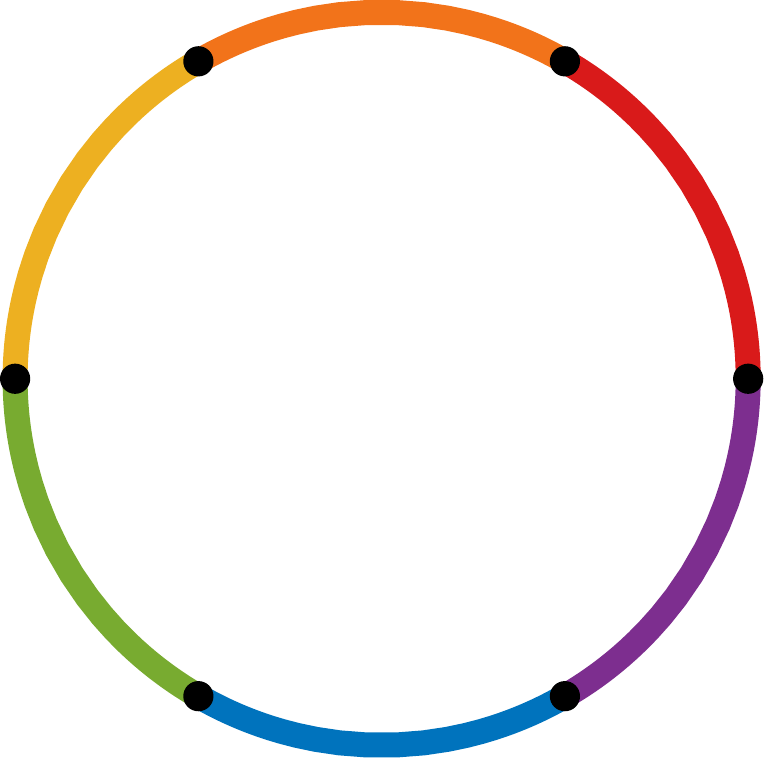}
    \caption{The solutions \eqref{gsol} for $N=3,4,5,6$, with $n=0,1,\dots, N-1$, $x_0 = 0$ and $\alpha=1$. We used $n=0$ (red), $1$ (orange), $2$ (yellow), $3$ (green), $4$ (blue), and $5$ (purple).}
    \label{fig2}
\end{figure}

\begin{figure}[!ht]
    \centering
    \includegraphics[scale=0.2]{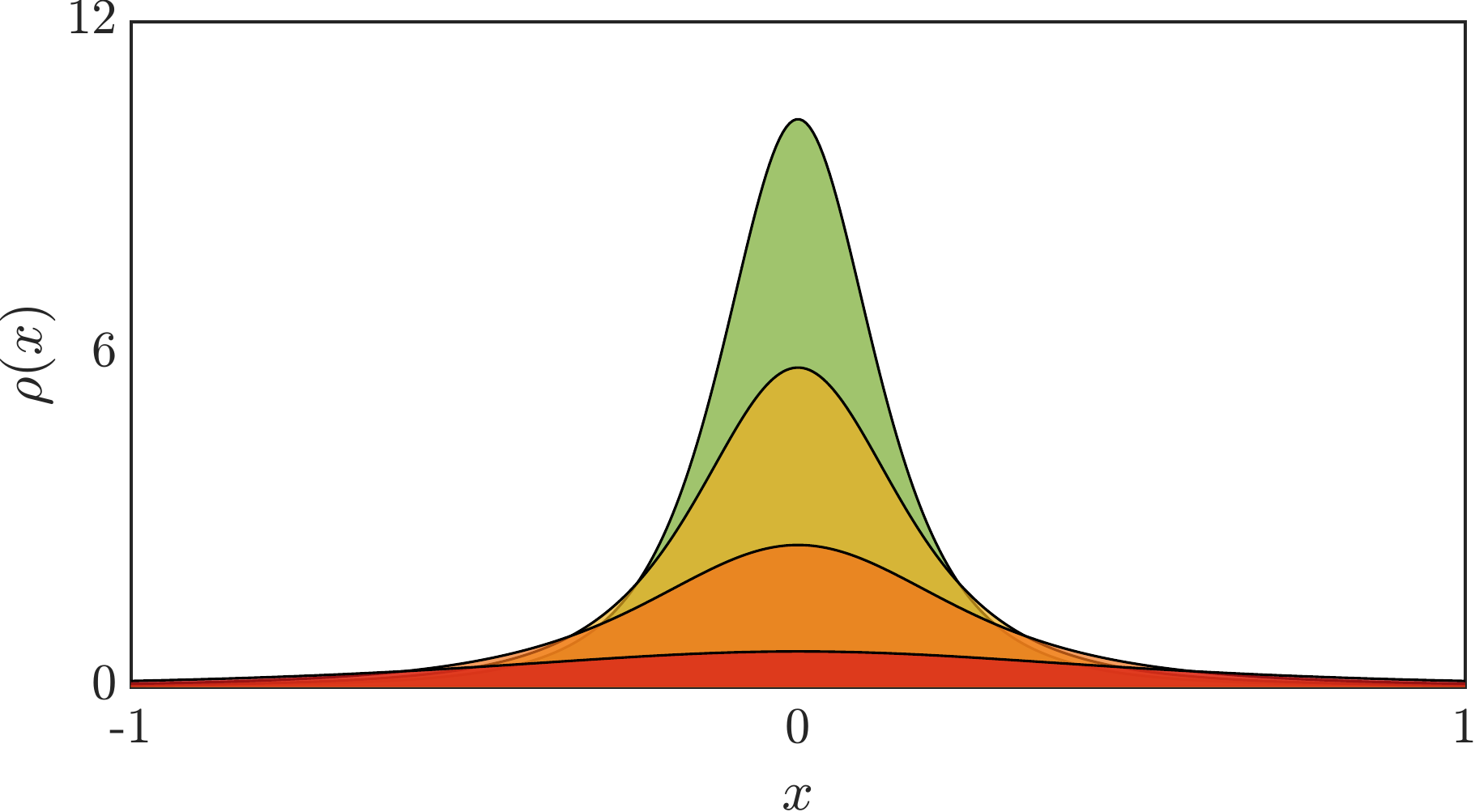}\\
    \includegraphics[scale=0.2]{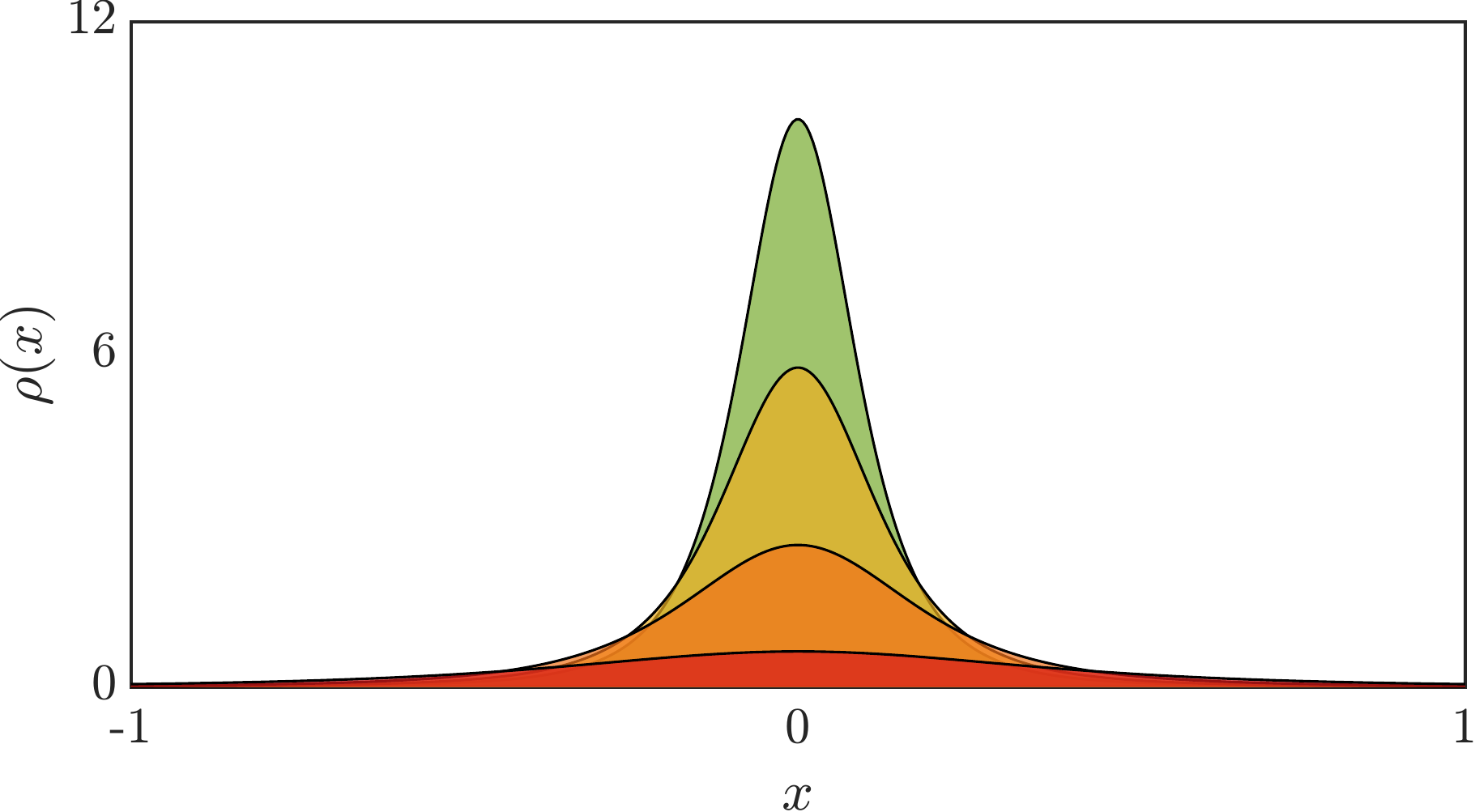}\\
    \includegraphics[scale=0.2]{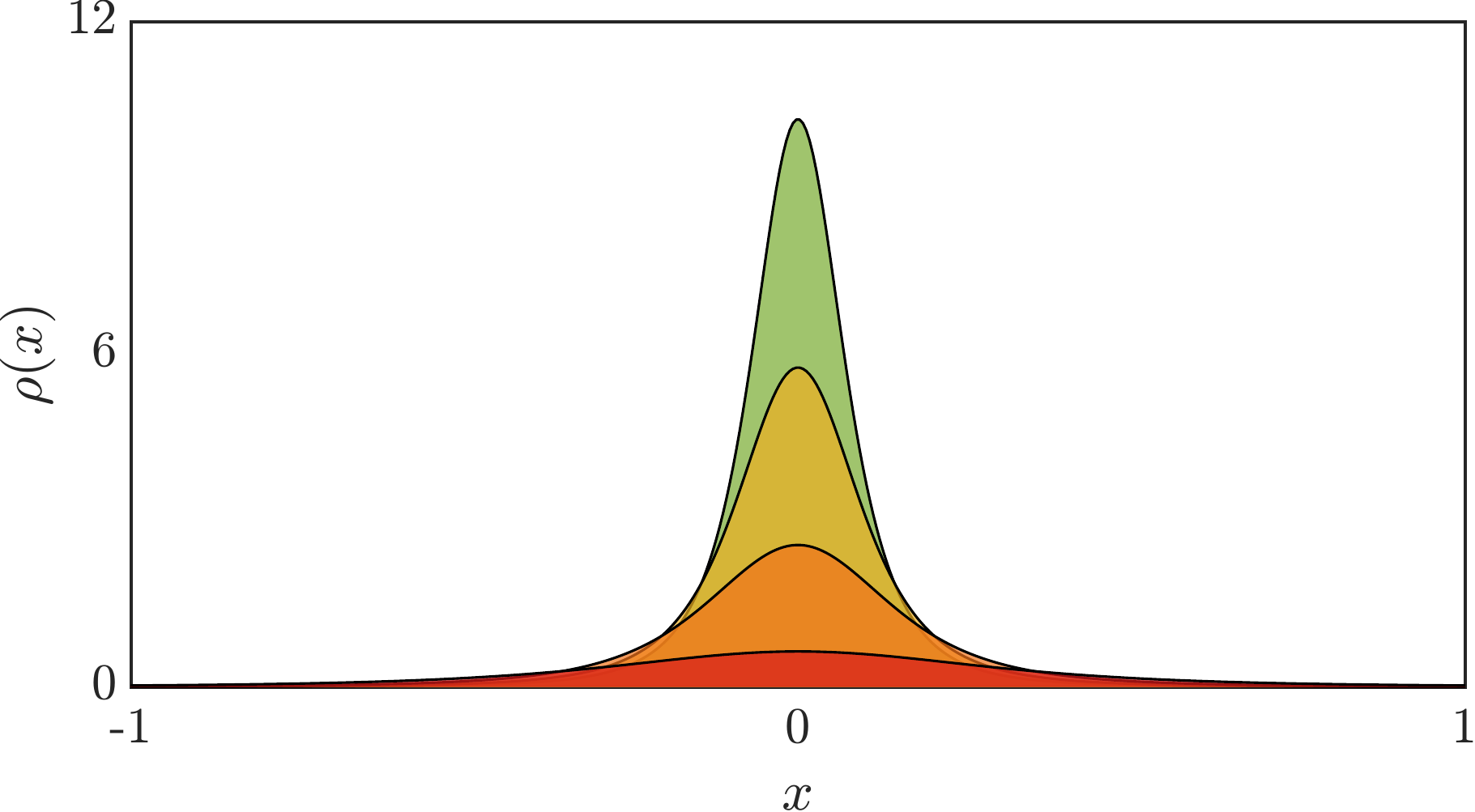}\\
    \includegraphics[scale=0.2]{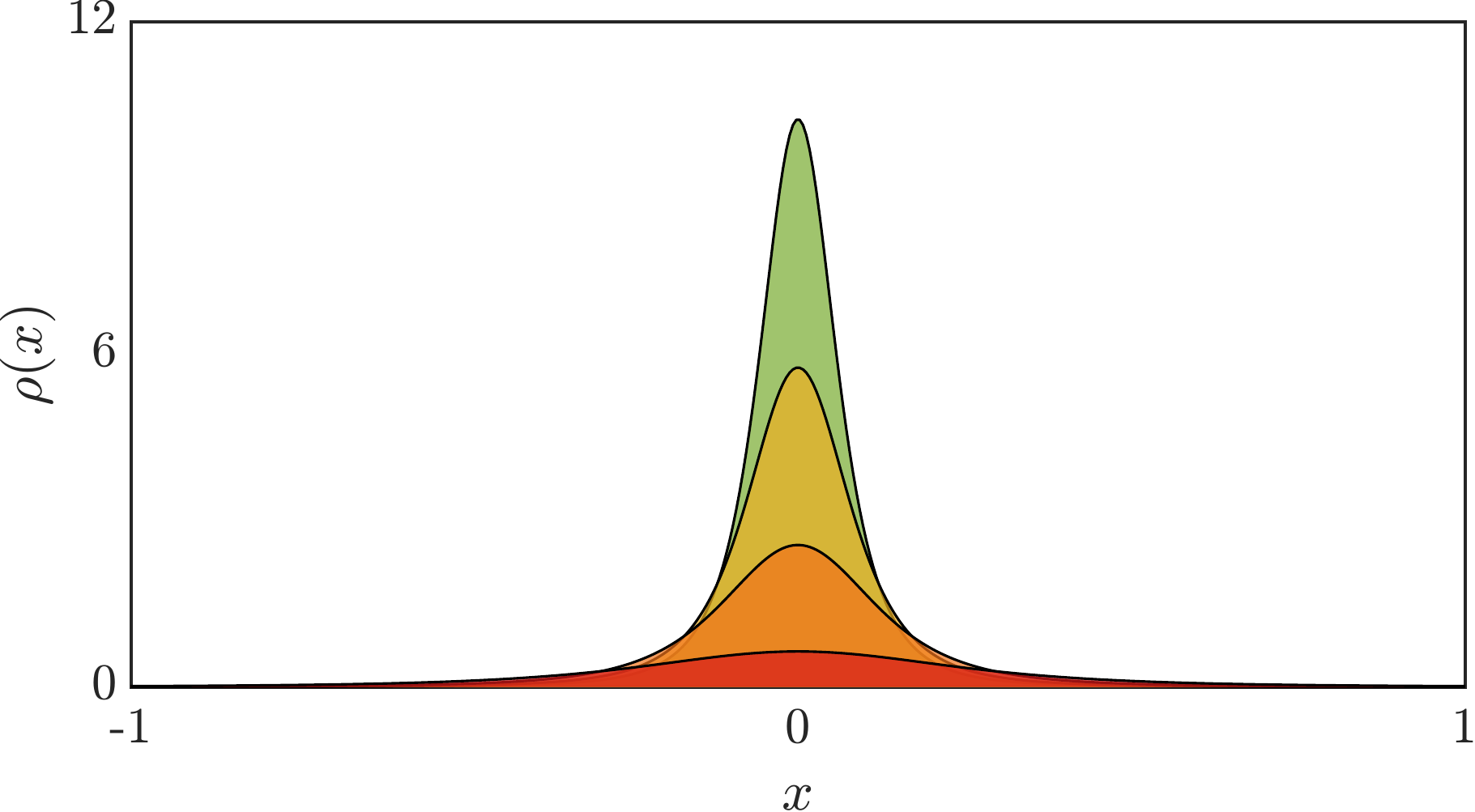}
    \caption{The energy density \eqref{rhox} for $N=3,4,5,6$, with $n=0$, $x_0=0$ and $\alpha=0.2$ (red), $0.4$ (orange), $0.6$ (yellow) and $0.8$ (green).}
    \label{fig3}
\end{figure}

The solution for the complex field is given by
\begin{equation}
\label{gsol}
    \varphi(x) = \exp\PC{i\PC{\frac{2}{N}\arccot\PC{2\alpha N\PC{x-x_0}} + \frac{2n\pi}{N}}},
\end{equation}
which continuously connects two adjacent global minima of the potential. It is depicted in Fig. \ref{fig2} for $N=3,4,5$ and $6$, following the left to right and the top to bottom sequence, respectively. 

The above solution \eqref{gsol} represents a family of exact solutions which are controlled by $N$ and $\alpha$. They all have the following energy density
\begin{equation}
\label{rhox}
    \rho(x) = \frac{16\alpha^2}{\PC{1+4\alpha^2N^2\PC{x-x_0}^2}^2},
\end{equation}
which is depicted in Fig. \ref{fig3} for some values of $N$ and $\alpha$. One notes that the total energy is given by $E_{B}=4\alpha\pi/N$, which depends on both $N$ and $\alpha$. It increases linearly with $\alpha$, indicating that the thinner the solution $\Theta(x)$, the higher the energy becomes, as expected. On the other hand, the energy diminishes as $N$ increases, vanishing in the limit $N\to\infty$. In this regime, the discrete $\mathbb{Z}_N$ symmetry of the model is promoted to the continuous $U(1)$ symmetry, making the discrete set of critical points given by \eqref{min} populate the unit circle, thereby forbidding the existence of solutions, in accordance with Eq. \eqref{theta}.

In summary, we have investigated a family of models described by a complex scalar field in $1+1$ spacetime dimensions, controlled by the discrete $Z_N$ symmetry. We have found a family of models that support exact solutions and expect that they may be of current interest to many areas, in particular, for building networks of localized structures; this was previously investigated in Ref. \cite{orb2}, and the family of models studied above may lead us to different networks. Here, an issue of interest is directly related to the fact that the tiling of the plane with regular polygons, can only be implemented with regular triangles, squares and regular hexagons, which correspond to the $Z_N$ symmetry with $N=3,4$ and $6$, so we may find obstructions for other values of $N$. In this context, in \cite{net1,net2,net3,net4,net5} the authors investigated junctions and the tiling of the plane with regular polygons, with and without supersymmetry. This is also of interest in condensed matter: for example, in Ref. \cite{rec1}, the study focuses on the construction of the ground state of a chiral magnet with square anisotropy. It is shown that the system may support domain wall networks as stable ground states, with the domain wall network turning out to become a skyrmion crystal; see also Ref. \cite{rec20,rec2,rec21,rec3}. The models investigated in the present work may also be used to study scalar dark matter in high energy physics, following the lines of Refs. \cite{DM1,DM2,DM3}, with the discrete $Z_N$ symmetry contributing to stabilize the scalar dark matter particles. 

A direct extension of the model would take into account the presence of another complex field, including interactions via the potential, in a way similar to the Higgs portal approach usually considered in high energy physics \cite{HP}. The presence of the second field may also be used to impose a geometric modification on the solution of the first field, changing the internal structure of the solution and opening up other possibilities of practical interest. This was previously considered in Ref. \cite{geo1}, motivated by the experimental result described in \cite{geom}, which studied a sample of magnetic material in the presence of a geometric constriction. Moreover, following the results of Ref. \cite{pulse}, which investigated the control of domain wall polarity by electric pulses, as well as the case of fermions in the presence of a localized bosonic structure previously considered in Ref. \cite{JR}, the study of fermions was further explored in \cite{geo2,geo3}. The new results revealed that the fermionic behavior can be significantly altered, inducing the presence of bands of states that contribute to modifying the electronic conduction in magnetic materials.

\vspace{0.4cm}
{{\bf Acknowledgments:} This work was partially financed by Coordenação de Aperfeiçoamento de Pessoal de Nível Superior (CAPES), Grant 88887.132514/2025-00 (GSS), and by Conselho Nacional de Desenvolvimento Científico e Tecnológico (CNPq), Grants 303469/2019-6 (DB), 402830/2023-7 (DB and RM), and 304344/2025-7 (RM).}

\vspace{0.4cm}

{\bf Data Availability Statement:} This manuscript has no associated data. [Author’s comment: Data sharing is not applicable to this article, as no datasets were generated or analyzed during the current study.]

\vspace{0.4cm}

{\bf Code Availability Statement:} This manuscript has no associated
code/software. [Author’s comment: Code/Software sharing is not applicable to this article as no code/software was generated or analyzed duringthe current study.]


\begin{thebibliography}{99}
\bibitem{F}P. Fendley, S.D. Mathur, C. Vafa, and N.P. Warner, Phys. Lett. B 243, 157 (1990).
\bibitem{AT}E.R.C. Abraham and P.K. Townsend, Nucl. Phys. B 351, 313 (1991).
\bibitem{B1}D. Bazeia, J. Menezes and M.M. Santos, Phys. Lett. B 521, 418 (2001).
\bibitem{B2}D. Bazeia, J. Menezes and M.M. Santos, Nucl. Phys. B 636, 132 (2002).
\bibitem{orb1}V.I. Afonso, D. Bazeia, M.A. González León, L. Losano and J. Mateos Guilarte, Phys. Lett. B 662, 75 (2008).
\bibitem{orb2}V.I. Afonso, D. Bazeia, M.A. González León, L. Losano and J. Mateos Guilarte, Nucl. Phys. B 810, 427 (2009).
\bibitem{orb3}A. Alonso-Izquierdo, M.A. González León, J. Martín Vaquero and M. de la Torre Mayado, Commun. Nonlinear Sci. Numer. Simul. 103, 106011 (2021).
\bibitem{DP}D. Bazeia, L. Losano, and J.M.C. Malbouisson, Phys. Rev. D 66, 101701(R) (2002).
\bibitem{rec20}S. Mühlbauer, B. Binz, F. Jonietz, et al. Science 323, 915
(2009).
\bibitem{rec2}X.Z. Yu, Y. Onose, N. Kanazawa, et al. 
Nature 465, 901 (2010).
\bibitem{rec21}S. Divic, H. Ling, T. Pereg-Barnea, and A. Paramekanti, 
Phys. Rev. B 105, 035156 (2022).
\bibitem{DM1}M. Frigerio, N. Grimbaum-Yamamoto, and T. Hambye, SciPost Phys. 15, 177 (2023).

\bibitem{bogo}E.B. Bogomol'nyi, Sov. J. Nucl. Phys. 24, 449 (1976).
\bibitem{PS}M.K. Prasad and C.M. Sommerfield, Phys. Rev. Lett. 35, 760 (1975).
\bibitem{net1}G.W. Gibbons and P.K. Townsend, Phys. Rev. Lett. 83, 1727 (1999).
\bibitem{net2}P.M. Saffin, Phys. Rev. Lett. 83, 4249 (1999).
\bibitem{net3}D. Bazeia and F.A. Brito, Phys. Rev. Lett. 84, 1094 (2000).
\bibitem{net4}S.M. Carroll, S. Hellerman, and M. Trodden, Phys. Rev. D 61, 065001 (2000).
\bibitem{net5}D. Bazeia and F.A. Brito, Phys. Rev. D 61, 105019 (2000). 
\bibitem{rec1}S. Lee, T. Fujimori, M. Nitta, and S.K. Kim, Phys. Rev. B 112, 064422 (2025).

\bibitem{rec3}Y. Amari and M. Nitta, Phys. Rev. B 111, 134441 (2025).
\bibitem{DM2}C.-W. Chiang and B.-Q. Lu, J. High Energy Phys. 07, 082 (2020).
\bibitem{DM3}D. Jurčiukonis and  L. Lavoura, 
Prog. Theor. Exp. Phys. 2023, 023B02 (2023). 
\bibitem{HP}G. Arcadi, A. Djouadi, M. Kado, Eur. Phys. J. C 81, 653 (2021).
\bibitem{geo1}D. Bazeia, M.A. Liao, and M.A. Marques, Eur. Phys. J. Plus 135, 383 (2020).
\bibitem{geom}P.-O. Jubert, R. Allenspach, and A. Bischof, Phys. Rev. B
69, 220410(R) (2004).
\bibitem{pulse}A. Vanhaverbeke, A. Bischof, and R. Allenspach, Phys. Rev.
Lett. 101, 107202 (2008).
\bibitem{JR}R. Jackiw and C. Rebbi, Phys. Rev. D 13, 3398 (1976).
\bibitem{geo2} D. Bazeia, A. Mohammadi, D. C. Moreira, Phys. Rev. D 103, 025003 (2021).
\bibitem{geo3}D. Bazeia and F.C. Simas, Eur. Phys. J. C 84, 1039L (2024). 

\end{thebibliography}
\end{document}